\documentstyle[12pt,epsf]{article}

\renewcommand{\thefootnote}{\fnsymbol{footnote}}

\begin{document}
\sloppy
\sloppy
\sloppy
$\ $
\begin{flushright}{UT-732,\ \ 1995}\end{flushright}
\vskip 1.5 truecm

\begin{center}
{\large{\bf Gribov Problem and BRST Symmetry}}{%
\renewcommand{\thefootnote}{\fnsymbol{footnote}}
\footnote[2]{To be published in the Proceedings of International Symposium on
{\bf BRS Symmetry} on the Occasion of its 20th Anniversary, September 18-22,
1995, RIMS, Kyoto University,  Kyoto, Japan(Universal Academy Press, Tokyo)}
}%
\end{center}
\vskip .75 truecm
\centerline{\bf Kazuo Fujikawa}
\vskip .4 truecm
\centerline {\it Department of Physics,University of Tokyo}
\centerline {\it Bunkyo-ku,Tokyo 113,Japan}
\vskip 1. truecm

\makeatletter
\@addtoreset{equation}{section}
\def\theequation{\thesection.\arabic{equation}}
\makeatother

\vskip 1. truecm

\begin{abstract}
After a brief historical comment on the study of BRS(or BRST) symmetry , we
 discuss the quantization  of gauge theories with Gribov copies.  A path
integral with BRST symmetry can be
formulated by summing the Gribov-type  copies in a very specific way if
the functional correspondence between $\tau$ and the gauge parameter
$\omega$ defined by $\tau (x) = f( A_{\mu}^{\omega}(x))$ is ``globally single
valued'', where $f( A_{\mu}^{\omega}(x)) = 0 $ specifies the gauge condition.
As an example of the theory which satisfies this criterion, we comment on a
 soluble gauge model with Gribov-type copies recently analyzed  by Friedberg,
Lee, Pang and Ren.
We also comment on a possible connection of the dynamical instability of BRST
symmetry with  the Gribov problem on the basis of an index notion.

\par

\end{abstract}

\newpage
\section{Introduction}

\par
The BRS(or BRST) symmetry and its applications to gauge field theories in
general have been developed and completed by the efforts of many authors. The
Faddeev-Popov ghost fields[1]  and the so-called Slavnov-Taylor identities[2]
paved a way to
the discovery of the BRST symmetry[3]. The Nakanishi-Lautrup $B$-field[4],
which was
introduced to define the Landau gauge in QED, played an important role to
define the off-shell closure of the BRST algebra. The BRST symmetry has
completely
changed the picture of perturbative renormalization of Yang-Mills fields[5].
The path integral formalism[6] as well as anti-field anti-bracket formalism[7]
extended the Dirac's general idea of the quantization of singular
Lagrangians[8] to the maximum extent. These developemtns have been summarized
in several
review articles[9] and a text book[10].

As for the study of BRST symmetry in Japan\footnote{Since many speakers of the
Symposium commented on some historical account of their encounter with the BRST
symmetry, this short account of mine  was added in this written version.} , the
first significant contribution to the subject
was made by Kugo and Ojima[11], who utilized the BRST charge to extract the
positive metric Hilbert space from an indefinite metric theory containing
Faddeev-Popov
ghost fields and the time component of gauge field. Their formalism completed a
previous attempt toward this direction[12]. The work of Kugo and Ojima was
immediately and highly appreciated by two leading Japanese theorists, K.
Nishijima and N. Nakanishi, both of whom applied the idea to the formal
analyses of
Einstein gravity[13].
It was then  pointed out[14] that the basic mechanism of selecting the positive
 metric space in the BRST approach is the BRST cohomology, although at that
time this terminology was not known: By using the notion of BRST superfield, a
simple and explicit classification of BRST invariant states was given.
The basic tone in Japan was an emphasis on the operator treatment of the BRST
symmetry rather than on the path integral analysis[15].

In the meantime, the path integral treatment of covariant string theory by
Polyakov appeared in 1981[16]. I applied the BRST symmetry to the string theory
partly as an application of the path integral treatment of anomalies[17]. The
Virasoro
condition[18] was then recognized as a result of BRST invariance. It was also
noted that the {\em proper part} of BRST symmetry is preserved for any (target)
space-time dimensions, which is related to  non-critical string theory.   The
ghost number anomaly in string theory
\begin{equation}
\partial_{\mu} j^{\mu}_{gh} =  - \frac{3}{4\pi}\sqrt{g}R
\end{equation}
was first encountered there[19]; this anomaly, which contains the scalar
curvature in the right-hand side,  was later recognized as a local
version of the Riemann-Roch theorem and as such it naturally played an
important
role in string theory and topological field theories. In the context of the
Gribov problem to be discussed below( in particular, the vanishing
Faddeev-Popov determinant), the ghost number anomaly in (1.1) generally shows
 the appearance of the zero modes of ghost fields; for example, 3 ghost zero
modes on a sphere. It is known that the appearance of those ghost zero modes
does not spoil the BRST symmetric formulation of string theory, at least in the
first quantization of strings. This is analogous to the zero modes of fermions
in the background of instantons. The mere appearance of zero modes of the
Faddeev-Popov determinant thus does not necessarily imply the difficulty of
BRST symmetric formulation. More about this will be commented on in Section 4.

About 6 months later since  my preprint on the BRST formulation of string
theory  appeared , Kato
and Ogawa , both of whom were graduate students at Kyoto University at that
time , performed a very detailed BRST operator analysis of the bosonic string
theory[20]. It is well known
that their paper greatly influenced the later developments in the covariant
treatment of strings and string field theories[21].
The BRS or BRST symmetry thus became a standard terminology  of a wider class
of particle theorists, as was noted by J. Gomis at this Symposium.

Coming to a fundamental aspect of BRST symmetry, I would like to comment on one
of the possible physical
interpretations of the famous factor $1/2$ in the BRST transformation law of
ghost fields\footnote{This issue was raised by R. Stora at the Symposium.}, on
the basis of the BRST invariance of the path integral measure. This property is
seen  particularly clearly if one looks at the gravitational BRST symmetry. The
structure ``constant'' $ F^{\mu}_{\alpha\beta}$ for the coordinate
transformation is defined  by
\begin{eqnarray}
[ \xi^{\rho}(x)\partial_{\rho}, \eta^{\lambda}(x)\partial_{\lambda} ]\phi (x)
&=&
(\xi^{\alpha}\partial_{\alpha}\eta^{\mu} -
\partial_{\beta}\xi^{\mu}\eta^{\beta})\partial_{\mu}\phi (x)\nonumber\\
&\equiv&  F^{\mu}_{\alpha\beta}\xi^{\alpha}\eta^{\beta}\partial_{\mu}\phi (x)
\end{eqnarray}
The BRST transformation of ghost fields is then defined with a factor of $1/2$
\begin{eqnarray}
c^{\mu} (x, \theta) &=& c^{\mu} (x) + i\theta \frac{1}{2}F^{\mu}_{\alpha\beta}
c^{\alpha}(x)c^{\beta}(x)\nonumber\\
&=& c^{\mu} (x) + i\theta c^{\rho}(x)\partial_{\rho}c^{\mu}(x)
\end{eqnarray}
We here used the superfield notation where the second component proportinal to
the Grassmann number $\theta$ stands for the BRST tranform of the first
component. The transformation law in (1.3) does not appear to have an explicit
connection with the general coordinate transformation, if one identifies
$i\theta c^{\rho}(x)$ with the parameter of coordinate transformation, as is
the case in
 metric variables. However, if one considers the differential of the superfield
in (1.3), one obtains
\begin{equation}
d c^{\mu} (x, \theta) = d c^{\mu} (x) + i\theta [c^{\rho}(x)\partial_{\rho}d
c^{\mu} (x) - \partial_{\rho}c^{\mu}(x)dc^{\rho}(x)]
\end{equation}
This expression, when regarded as a transformation law of $dc^{\mu}(x)$,  has
the structure of general coordinate transformation.  Without the factor of
$1/2$, the differential (1.4) would not have the form of coordinate
transformation.
It is
shown that this last structure, when combined with BRST transformation law of
metric variables, is crucial to establish the BRST invariance of the
gravitational path integral measure without having an artificial anomaly( or
Jacobian)[22]. Note that the Jacobian for the integral measure is defined by
(1.4).

Although the success of the BRST symmetry is impressive in the realm of
perturbation theory of gauge fields in general, the future prospect of the BRST
symmetry in connection with non-perturbative analyses is not quite bright, as
was emphasized by C. Becchi at this Symposium[23]. This is due to the presence
of the so-called Gribov problem[24]$\sim$[27] and the lack of convincing
and simple prescription to deal with it in the framework of BRST symmetry. In
the next Section I would like to discuss this issue in some detail.

\section{BRST invariant path integral in the presence of Gribov copies}
\par
There have been many attempts to provide a simple prescription to quantize the
theory with Gribov problems. As one of these attempts to quantize the theory
with Gribov copies by using BRST symmetry[28][29], we here recapitulate the
essence of the argument presented in [28]. We start with the Faddeev-Popov
formulation of the Feynman-type gauge condition [30]. The vacuum-to-vacuum
transition amplitude is defined by
\begin{equation}
\langle +\infty|-\infty\rangle = {\int}{\cal D}A_{\mu}^{\omega}{\cal D}C
\delta(\partial^{\mu}A_{\mu}^{\omega} - C)\Delta(A)
exp\{iS(A_{\mu}^{\omega}) - \frac{i}{2\alpha}{\int}C(x)^{2}dx\}
\end{equation}
where $S(A_{\mu}^{\omega})$ stands for the action invariant under the
Yang-Mills local gauge transformation. The positive constant $\alpha$  is a
gauge fixing parameter which specifies the Feynman-type gauge condition.
In the following we often suppress the internal symmetry indices, and instead
we write the gauge parameter explicitly: $A_{\mu}^{\omega}$ indicates the gauge
field which is obtained from $A_{\mu}$ by a gauge transformation specified by
$\omega(x)$. The determinant factor $\Delta(A)$ is defined by[1][30]

\begin{eqnarray}
\Delta(A)^{-1}&=&{\int}{\cal D}\omega{\cal D}C\
\delta(\partial^{\mu}A_{\mu}^{\omega} - C)
exp\{ - \frac{i}{2\alpha}{\int}C(x)^{2}dx\}\nonumber\\
&\approx&const\{\sum_{k}|det[\frac{\partial}{\partial\omega_{k}}
\partial^{\mu}A_{\mu}^{{\omega}_{k}}]|^{-1}\}
\end{eqnarray}
where the summation runs over all the gauge equivalent configurations
satisfying $\partial^{\mu}A_{\mu}^{{\omega}_{k}}=0$ , which were
found by Gribov[24] and others[25]$\sim$[27].
Equation(2.2) is valid only for sufficiently small $\alpha$, since
the parameter $\omega^{\prime}(\omega,A,C)$ defined by
\begin{equation}
\partial^{\mu}A_{\mu}^{\omega^{\prime}(\omega,A,C)}=
\partial^{\mu}A_{\mu}^{\omega}- C
\end{equation}
has a complicated branch structure for large $C$ in the presence of Gribov
ambiguities. Obviously the Feynman-type gauge formulation becomes even more
involved than the Landau-type gauge condition.

It was suggested in [28] to replace equation (2.1) by
\begin{equation}
\langle +\infty|-\infty\rangle = \frac{1}{N}{\int}{\cal D}A_{\mu}^{\omega}{\cal
D}C
\delta(\partial^{\mu}A_{\mu}^{\omega} -
C)det[\frac{\partial}{\partial\omega}\partial^{\mu}A_{\mu}^{\omega}]
exp\{iS(A_{\mu}^{\omega}) - \frac{i}{2\alpha}{\int}C(x)^{2}dx\}
\end{equation}
The crucial difference between (2.1) and (2.4) is that (2.4) is local in the
gauge space $\omega(x)$ (i.e., the gauge fixing factor and the compensating
factor are defined at the identical $\omega$ ),whereas $\Delta(A)$ in (2.1) is
gauge independent and involves a non-local factor in $\omega$ as is shown in
(2.2). As the determinant in (2.4) depends on $A_{\mu}^{\omega}$, the entire
integrand in (2.4) is in general no more degenerate with respect to gauge
equivalent configurations even if the gauge fixing term itself may be
degenerate for certain configurations. Another important point is that one
takes the absolute values of determinant factors in (2.2) thanks to the
definition of the $\delta$-function, whereas just the determinant which can be
negative as well as positive appears in (2.4).

It is easy to see that
(2.4) can be rewritten as
\begin{equation}
\langle +\infty|-\infty\rangle = \frac{1}{\tilde{N}}{\int}{\cal
D}A_{\mu}^{\omega}{\cal D}B{\cal D}\bar{c}{\cal D}c
\ exp\{iS(A_{\mu}^{\omega}) +i {\int}{\cal L}_{g}dx\}
\end{equation}
where
\begin{equation}
{\cal L}_{g}= -\partial^{\mu}B^{a}A_{\mu}^{a\omega} + i\partial^{\mu}
\bar{c}^{a}(\partial_{\mu}-gf^{abc}A_{\mu}^{b\omega})c^{c}
+ \frac{\alpha}{2}B^{a}B^{a}
\end{equation}
with $B^{a}$ the Lagrangian multiplier field, and $\bar{c}^{a}$ and
$c^{a}$ the (hermitian) Faddeev-Popov ghost fields; $f^{abc}$ is the
structure constant of the gauge group and $g$ is the gauge coupling constant.
If one imposes the hermiticity of $\bar{c}^{a}$ and $c^{a}$, the phase factor
of the determinant in (2.4) cannot be removed.
The normalization constant $\tilde{N}$  in (2.5)
includes the effect of Gaussian integral over $B$ in addition to $N$ in (2.4),
and in fact $\tilde{N}$ is independent of $\alpha$. See eq. (2.10).
This ${\cal L}_{g}$ as well as the starting gauge invariant Lagrangian are
invariant under the BRST transformation defined by
\begin{eqnarray}
\delta_{\theta} A_{\mu}^{a\omega} &=& i\theta [\partial_{\mu}c^{a} -gf^{abc}
A_{\mu}^{b\omega}c^{c}]\nonumber\\
\delta_{\theta} c^{a} &=& i\theta (g/2)f^{abc}c^{b}c^{c}\nonumber\\
\delta_{\theta} \bar{c}^{a} &=& \theta B^{a}\nonumber\\
\delta_{\theta} B^{a} &=& 0
\end{eqnarray}
where $\theta$  and  the ghost variables $c^{a}(x)$ and $\bar{c}^{a}(x)$ are
elements of the Grassmann algebra , i.e., $\theta^{2} = 0$. This transformation
can be confirmed to be nil-potent
$\delta^{2} = 0$, for example,
\begin{equation}
\delta_{\theta_{2}}(\delta_{\theta_{1}} A_{\mu}^{a\omega}) = 0
\end{equation}
One can also confirm that the path integral measure in (2.5) is invariant under
(2.7).
Note that the transformation (2.7) is ``local'' in the $\omega$ parameter;\
precisely for this property, the prescription in (2.4) was chosen.

To interprete  the path integral measure in (2.4) as the path integral  over
all the gauge field configurations divided by the gauge volume, namely
\begin{equation}
\langle +\infty|-\infty\rangle ={\int}\frac{{\cal D}A_{\mu}^{\omega}}{gauge\
volume(\omega)}
\ exp\{iS(A_{\mu}^{\omega})\}
\end{equation}
one needs to define the normalization factor in (2.4) by
\begin{eqnarray}
N &=& {\int}{\cal D}\omega{\cal D}C
\delta(\partial^{\mu}A_{\mu}^{\omega} -
C)det[\frac{\partial}{\partial\omega}\partial^{\mu}A_{\mu}^{\omega}]
exp\{- \frac{i}{2\alpha}{\int}C(x)^{2}dx\}\nonumber\\
  &=& {\int}{\cal D}\omega
det[\frac{\partial}{\partial\omega}\partial^{\mu}A_{\mu}^{\omega}]
exp\{-
\frac{i}{2\alpha}{\int}(\partial^{\mu}A_{\mu}^{\omega})^{2}dx\}\nonumber\\
  &=& {\int}{\cal D}\tau\ exp\{- \frac{i}{2\alpha}{\int}\tau (x)^{2}dx\}
\end{eqnarray}
where the function $\tau$ is defined by
\begin{equation}
\tau (x)\equiv \partial^{\mu}A_{\mu}^{\omega}(x)
\end{equation}
and the determinant factor is regarded as a Jacobian for the change
of variables from $\omega (x)$ to $\tau (x)$.
Although we use the Feynman-type gauge fixing (2.11) as a typical
example in this Section, one may replace (2.11) by
\begin{equation}
\tau (x)\equiv f( A_{\mu}^{\omega}(x))
\end{equation}
to deal with a more general gauge condition
\begin{equation}
 f( A_{\mu}^{\omega}(x)) = 0
\end{equation}
It is crucial to establish that the normalization factor in (2.10) is
independent of $A_{\mu}$. Only in this case, (2.4) defines an acceptable vacuum
transition amplitude. The Gribov ambiguity in the present
case appears as a non-unique correspondence between $\tau (x)$ and $\omega (x)$
in (2.11), as is schematically shown in Fig. 1 which includes 3 Gribov copies.
The path
integral in (2.10) is performed along the contour in Fig. 1. As the
Gaussian function is regular at any finite point, the complicated
contour in Fig. 1 gives rise to the same result in (2.10) as a
 contour corresponding to $A_{\mu} = 0$. In the present path integral
formulation, the evaluation of the normalization factor in (2.10) is the only
place where we explicitly encounter the multiple solutions of gauge fixing
condition.[ If the normalization factor $N$ should depend on gauge field
$A_{\mu}$, the factor $N$, which is gauge independent in the sense that we
integrated over entire gauge orbit, needs
to be taken inside the path integral in (2.4). In this case one looses the
simplicity of the formula (2.4).]

The basic assumption we have to make is therefore that (2.11) in the context of
the path integral (2.10) is ``globally single-valued'', in the sense that the
asymptotic functional correspondence between $\omega$ and $\tau$ is little
affected by a fixed $A_{\mu}$ with
 $\partial^{\mu}A_{\mu} = 0$ [28]; \ Fig.1 satisfies this requirement.
This assumption appears to be physically reasonable if the second derivative
term of the gauge orbit parameter dominates the functional
correspondence in (2.11),
though it has not been established mathematically. To define the functional
correspondence between $\omega$ and $\tau$ in (2.11), one needs in
general some notion of norm such as $L^{2}$-norm for which the Coulomb gauge
vacuum
is unique [26][27].  The functional configurations which are square integrable
however have zero measure in the path integral[31], and this makes the precise
analysis of (2.11) very complicated:\  At least what we need to do is to start
with an expansion of a generic
field variable in terms of some complete orthonormal basis set (which means
that the field is  inside the $L^{2}$-space)  and then let each expansion
coefficient vary from
$-\infty$ to $+\infty$(which means that the field is outside the
$L^{2}$-space).

 The indefinite signature of the determinant factor in (2.4) is not a
difficulty in the framework of indefinite metric field theory [11]$\sim$[15]
since  the determinant factor is associated with the Faddeev-Popov ghost fields
and the BRST cohomology selects the positive definite physical space.  On the
other hand , the Gribov problem may aslo suggest  that one cannot bring the
relation(2.11) with fixed $A_{\mu}$ to $\partial^{\mu}A_{\mu}^{\omega}=0$ by
any gauge transformation [25]. If this is the case, the asymptotic behavior of
the mapping (2.11) is in general modified by $A_{\mu}$ and our prescription
cannot be justified.

It is not known at this moment to what extent the global single-valuedness of
(2.11) in the sense of the path integral (2.10) is satisfied by 4-dimensional
Yang-Mills fields. It is therefore very important to deepen our understanding
of this problem by studying a simpler model with a finite number of degrees of
freedom. Recently, a very detailed analysis of a soluble gauge model with
Gribov-type copies has been performed by Friedberg, Lee, Pang, and Ren[32]. As
is
explained below, their model nicely satisfies our criterion in (2.11).
Although it is not clear whether the properties exhibited by this soluble
model are generic for theories with Gribov copies, it is expected that the
soluble model will help increase our understanding of the problem and may
provide
us  a deeper insight into the possible quantization of gauge theories.

\section{A soluble gauge model with Gribov-type copies}
{\bf  3.1, THE MODEL OF FRIEDBERG, LEE, PANG, AND REN }
\par
The soluble gauge model of Friedberg, Lee, Pang and Ren[32] is defined by
\begin{equation}
{\cal L}= \frac{1}{2}\{[\dot{X}(t)+g\xi(t)Y(t)]^{2} + [\dot{Y}(t) -
g\xi(t)X(t)]^{2} + [\dot{Z}(t) - \xi(t)]^{2}\} - U(X(t)^{2} +Y(t)^{2})
\end{equation}
where $\dot{X}(t)$, for example, means the time derivative of $X(t)$, and the
potential $U$ depends only on the combination $X^{2} + Y^{2}$. This Lagrangian
is invariant under a local gauge transformation
parametrized by $\omega (t)$ ,
\begin{eqnarray}
X^{\omega}(t) &=& X(t)\cos g\omega(t) - Y(t)\sin g\omega(t)\nonumber\\
Y^{\omega}(t) &=& X(t)\sin g\omega(t) + Y(t)\cos g\omega(t)\nonumber\\
Z^{\omega}(t) &=& Z(t) + \omega(t)\nonumber\\
\xi^{\omega}(t) &=& \xi(t) + \dot{\omega}(t)
\end{eqnarray}

The gauge condition ( an analogue of $A_{0} = 0$ gauge )
\begin{equation}
\xi(t) = 0
\end{equation}
or ( an analogue of $A_{3} = 0$ gauge )
\begin{equation}
Z(t) = 0
\end{equation}
is well-defined without suffering from Gribov-type copies. However,
it was shown in [32] that the gauge condition ( an analogue of the Coulomb
gauge )
\begin{equation}
Z(t) - \lambda X(t) = 0
\end{equation}
with a constant $\lambda$ suffers from the Gribov- type complications. This is
seen by using the notation in (3.2) as
\begin{eqnarray}
Z^{\omega}(t) - \lambda X^{\omega}(t) &=& Z(t) + \omega(t) - \lambda
X(t)\cos g\omega(t) + \lambda Y(t)\sin g\omega(t)\nonumber\\
&=& \omega(t) +\lambda\sqrt{X^{2} +Y^{2}}[\cos \phi(t) -\cos (g\omega(t) +
\phi(t))] = 0
\end{eqnarray}
where we used the relation(3.5) and
\begin{equation}
X(t) = \sqrt{X^{2} +Y^{2}}\cos\phi(t), \ \ Y(t) =\sqrt{X^{2} +Y^{2}}\sin\phi(t)
\end{equation}
{}From a view point of gauge fixing, $\omega(t)= 0$ is a solution of
(3.6) if (3.5) is satisfied. By analyzing the crossing points of two
graphs in $(\omega, \eta) $ plane defined by
\begin{eqnarray}
\eta &=& \frac{1}{\lambda\sqrt{X^{2}+Y^{2}}}\omega\nonumber\\
\eta &=& \cos ( g\omega +\phi) -\cos\phi
\end{eqnarray}
one can confirm that eq.(3.6) in general has more than one solutions for
$\omega$.

{}From a view point of general gauge fixing procedure, we here regard the
algebraic gauge fixing such as (3.3) and (3.4) well-defined; \ in the analysis
of the Gribov problem in Ref.[25], the algebraic gauge fixing [26] is excluded.

The authors in Ref.[32] started with the Hamiltonian formulated
in terms of  the well-defined gauge $\xi (t) = 0 $ in (3.3) and then
faithfully rewrote the Hamiltonian in terms of the variables defined by the
``Coulomb gauge'' in (3.5). By this way, the authors in [32] analyzed in detail
the problem related to the Gribov copies and the so-called Gribov horizons
where the Faddeev-Popov determinant
vanishes. They thus arrived at a prescription which sums over all the
Gribov-type copies in a very specific way. As is clear from their derivation,
their specification satisfies the unitarity and gauge independence.

In the context of  BRST invariant path integral discussed in Section 2, the
crucial relation (2.11) becomes
\begin{eqnarray}
\tau (t) &=& Z^{\omega}(t) - \lambda X^{\omega}(t)\nonumber\\
         &=& \omega(t) + Z(t) -\lambda X(t)\cos g\omega(t)  +\lambda Y(t)\sin
g\omega(t)
\end{eqnarray}
in the present model. For $X = Y =0$ , the functional correspondence between
$\omega$ and $\tau$   is one-to-one and monotonous for any
fixed value of $t$ . When one varies $X(t), Y(t)$ and $Z(t)$ continuously, one
deforms this monotonous curve continuously. But the asymptotic correspondence
between $\omega (t)$ and $\tau (t)$ at $\omega(t) = \pm\infty$ for each value
of $t$ is still kept preserved, at
least for any fixed $X(t), Y(t)$ and $Z(t)$ . This correspondence
between $\omega (t)$ and  $\tau (t)$ thus satisfies our criterion
discussed in connection with (2.11). The absence of terms which contain the
derivatives of $\omega (t)$ in (3.9) makes the functional correspondence in
(3.9) well-defined and transparent.

{}From a view point of gauge fixing in (3.6), this ``globally single-valued''
correspondence between $\omega$ and $\tau$ means that one always obtains an
{\em odd} number of solutions for (3.6).
The prescription in [32] is then viewed as a sum of all these solutions with
signature factors specified by the signature of the Faddeev-Popov determinant
\begin{equation}
det\{\frac{\partial}{\partial\omega (t^{\prime})}[Z^{\omega}(t)-\lambda
X^{\omega}(t)]\} = det\{[1 + \lambda g Y^{\omega}(t)]
\delta(t-t^{\prime})\}
\end{equation}
evaluated at the point of solutions , $\omega = \omega(\sqrt{X^{2}+Y^{2}},
\phi)$, of (3.6). The row and column indices of the matrix in (3.10) are
specified by $t$ and $t^{\prime}$, respectively.
In the context of BRST invariant formulation, a pair-wise
cancellation of Gribov-type  copies takes place,
except for one solution, in the calculation of the normalization
factor in (2.10) or (3.21) below.

\noindent
{\bf 3.2, BRST INVARIANT PATH INTEGRAL}

The relation (3.9) satisfies our criterion discussed in connection with (2.11).
We can thus define an analogue of (2.5) for the Lagrangian (3.1) by
\begin{equation}
\langle +\infty|-\infty\rangle = \frac{1}{\tilde{N}}{\int}d\mu
\ exp\{iS(X^{\omega},Y^{\omega},Z^{\omega},\xi^{\omega}) +i {\int}{\cal
L}_{g}dt\}
\end{equation}
where
\begin{eqnarray}
S(X^{\omega},Y^{\omega},Z^{\omega},\xi^{\omega}) &=& {\int}{\cal
L}(X^{\omega},Y^{\omega},Z^{\omega},\xi^{\omega}) dt\nonumber\\
 &=& S(X,Y,Z,\xi)
\end{eqnarray}
in terms of the Lagrangian ${\cal L}$ in (3.1). The gauge fixing part of (3.11)
is defined by
\begin{equation}
{\cal L}_{g} = -\beta \dot{B} \xi^{\omega} + B(Z^{\omega} - \lambda X^{\omega})
+ \beta i \dot{\bar{c}}\dot{c} -i\bar{c}( 1 + g\lambda Y^{\omega})c
+\frac{\alpha}{2}B^{2}
\end{equation}
where $\alpha,\beta$ and $\lambda$ are numerical constants, and $\bar{c}$ and
$c$  are (hermitian) Faddeev-Popov ghost fields. $B$ is a Lagrangian multiplier
field. Note that ${\cal L}_{g}$ is hermitian. The integral measure in (3.11) is
given by
\begin{equation}
d \mu = {\cal D}X^{\omega}{\cal D}Y^{\omega}{\cal D}Z^{\omega}{\cal
D}\xi^{\omega}{\cal D}B{\cal D}\bar{c}{\cal D}c
\end{equation}
The Lagrangians ${\cal L}$ and ${\cal L}_{g}$ and the path integral measure
(3.14) are invariant under the BRST transformation defined by
\begin{eqnarray}
X^{\omega}(t,\theta)&=& X^{\omega}(t) -i\theta g c(t) Y^{\omega}(t)\nonumber\\
Y^{\omega}(t,\theta)&=& Y^{\omega}(t) +i\theta g c(t) X^{\omega}(t)\nonumber\\
Z^{\omega}(t,\theta)&=& Z^{\omega}(t) +i\theta c(t)\nonumber\\
\xi^{\omega}(t,\theta)&=& \xi^{\omega}(t) +i\theta \dot{c}(t)\nonumber\\
c(t,\theta) &=& c(t)\nonumber\\
\bar{c}(t,\theta) &=& \bar{c}(t) + \theta B(t)
\end{eqnarray}
where the parameter $\theta$ is a Grassmann number, $\theta^{2} = 0$.
Note that $\theta$ and ghost variables anti-commute.
In (3.15) we used a BRST superfield notation: \ In this notation, the second
component of a superfield proportional to $\theta$ stands for the BRST
transformed field of the first component. The second component is invariant
under BRST transformation which ensures the nil-potency of the BRST charge. In
the operator notation to be defined later, one can write , for example,
\begin{equation}
X^{\omega}(t,\theta) = e^{-\theta Q} X^{\omega}(t,0) e^{\theta Q}
\end{equation}
with a nil-potent BRST charge $Q$, $\{ Q,Q\}_{+} = 0$. Namely, the BRST
transformation is a translation in $\theta$-space, and $\theta Q$ is analogous
to momentum operator.

In (3.11)$\sim$ (3.15), we explicitly wrote the gauge parameter
$\omega$
to emphasize that BRST transformation is ``local'' in the $\omega$-space. For
${\cal L}_{g}$ in (3.13), the relation (3.9) is replaced by ( an analogue of
the Landau gauge )
\begin{eqnarray}
\tau (t) &\equiv& \beta \dot{\xi}^{\omega}(t) + Z^{\omega}(t) - \lambda
X^{\omega}(t)\nonumber\\
&=& \beta \ddot\omega (t) + \omega(t) + \beta\dot{\xi}(t)
+ Z(t) -\lambda X(t)\cos g\omega(t)  +\lambda Y(t)\sin g\omega(t)
\end{eqnarray}
 The functional correspondence between $\omega$ and $\tau$ is monotonous and
one-to-one for weak $\xi(t), Z(t), X(t) $ and $Y(t)$ fields;
\ this is understood if one rewrites the relation (3.17) for Euclidean time $t
= - it_{E}$ by neglecting weak fields as
\begin{displaymath}
\tau (t_{E}) = ( - \beta\frac{d^{2}}{dt_{E}^{2}} + 1 ) \omega(t_{E})
\end{displaymath}
The Fourier transform of this relation gives a one-to-one monotonous
correspondence between the Fourier coefficients of $\tau$ and $\omega$ for
non-negative $\beta$. The asymptotic functional correspondence between $\omega$
and $\tau$ for weak field cases is preserved even for any fixed strong fields
$\xi(t), Z(t), X(t) $ and $Y(t)$ for non-negative $\beta$ to the extent that
the term linear in $\omega (t)$
dominates the cosine and sine terms.  The correspondence between $\tau$ and
$\omega$ in (3.17) is quite complicated for finite $\omega (t)$ due to the
presence of the derivatives of $\omega(t)$.

Thus (3.17) satisfies our criterion of BRST invariant path integral for any
non-negative $\beta$.
The relation (3.9) is recovered if one sets $\beta = 0$ in (3.17);\ the
non-zero parameter $\beta \neq 0$ however renders a canonical structure of the
theory
better-defined. For example, the kinetic term for ghost fields in (3.13)
disappears for $\beta = 0$. In this respect the gauge (3.5) is
also analogous to the unitary gauge.
In the following we set $\beta = \alpha > 0$ in (3.13),
\begin{equation}
{\cal L}_{g} = -\alpha \dot{B} \xi^{\omega} + B(Z^{\omega} - \lambda
X^{\omega})
+ \alpha i \dot{\bar{c}}\dot{c} -i\bar{c}( 1 + g\lambda Y^{\omega})c
+\frac{\alpha}{2}B^{2}
\end{equation}
and let $\alpha \rightarrow 0$ later. In the limit $\alpha = 0$, one recovers
the gauge condition (3.5)  defined in Ref.[32].
This procedure is analogous to $R_{\xi}$-gauge ( or the $\xi$-limiting
process of Lee and Yang [33] ), where the (singular) unitary
gauge is defined in the vanishing limit of the gauge parameter, $\xi
\rightarrow 0$:\ In (3.18) the parameter $\alpha$ plays the role of $\xi$ in
$R_{\xi}$-
gauge.

By using the BRST invariance, one can show the $\lambda$- independence
of (3.11) as follows:
\begin{equation}
\langle +\infty|-\infty\rangle_{\lambda+\delta \lambda} =
\langle +\infty|-\infty\rangle_{\lambda } -
\delta\lambda \frac{1}{\tilde{N}}{\int}d\mu [B(t)X^{\omega}(t) +
ig\bar{c}(t)Y^{\omega}(t)c(t)]
\ exp\{i {\int}({\cal L}+{\cal L}_{g})dt\}
\end{equation}
where we perturbatively expanded in the variation of ${\cal L}_{g}$  for a
change of the parameter $\lambda +\delta\lambda$ ,
\begin{equation}
{\cal L}_{g}(\lambda +\delta\lambda) = {\cal L}_{g}(\lambda)
- \delta\lambda [B(t)X^{\omega}(t) + ig\bar{c}(t)Y^{\omega}(t)c(t)]
\end{equation}
This expansion is justified since the normalization factor defined by (see
eq.(2.10))
\begin{eqnarray}
N &=& {\int}{\cal D}\tau\ exp\{- \frac{i}{2\alpha}{\int}\tau (x)^{2}dt\}
\end{eqnarray}
is independent of $\lambda$ provided that the global single-valuedness in
(3.17) is satisfied.  As was noted before, this path integral for $N$, which
depends on $\alpha$, is the only place where we explicitly encounter the
Gribov-type copies in the present approach.
By denoting the BRST transformed variables by prime, for example,
\begin{equation}
X^{\omega}(t)^{\prime} = X^{\omega}(t) -
ig\theta c(t)Y^{\omega}(t)
\end{equation}
we have a BRST identity ( or Slavnov-Taylor identity [2])
\begin{eqnarray}
\lefteqn{\frac{1}{\tilde{N}}{\int}d\mu \bar{c}(t)X^{\omega}(t)
\ exp\{i {\int}({\cal L}+{\cal L}_{g})dt\}}\nonumber\\
&=&
\frac{1}{\tilde{N}}{\int}d\mu^{\prime}
\bar{c}(t)^{\prime}X^{\omega}(t)^{\prime}
\ exp\{i {\int}({\cal L}^{\prime}+{\cal L}_{g}^{\prime})dt\}\nonumber\\
&=&
\frac{1}{\tilde{N}}{\int}d\mu \bar{c}(t)X^{\omega}(t)
\ exp\{i {\int}({\cal L}+{\cal L}_{g})dt\}\nonumber\\
&+&
\theta\frac{1}{\tilde{N}}{\int}d\mu [B(t)X^{\omega}(t) +
ig\bar{c}(t)Y^{\omega}(t)c(t)]
\ exp\{i {\int}({\cal L}+{\cal L}_{g})dt\}
\end{eqnarray}
where the first equality holds since the path integral is independent of the
naming of integration variables provided that the asymptotic
behavior and the boundary conditions are not modified by the change
of variables.  The second equality in (3.23) holds due to the BRST
invariance of the measure and the action
\begin{eqnarray}
d\mu^{\prime} &=& d\mu,\nonumber\\
{\cal L}^{\prime} +{\cal L}_{g}^{\prime} &=& {\cal L} + {\cal L}_{g}
\end{eqnarray}
but
\begin{equation}
\bar{c}(t)^{\prime}X^{\omega}(t)^{\prime}
=\bar{c}(t)X^{\omega}(t)
+\theta [B(t)X^{\omega}(t) + ig\bar{c}(t)Y^{\omega}(t)c(t)]
\end{equation}
{}From (3.23) one concludes
\begin{equation}
\frac{1}{\tilde{N}}{\int}d\mu [B(t)X^{\omega}(t) +
ig\bar{c}(t)Y^{\omega}(t)c(t)]
\ exp\{i {\int}({\cal L}+{\cal L}_{g})dt\} =0
\end{equation}
and thus
\begin{equation}
\langle +\infty|-\infty\rangle_{\lambda+\delta \lambda} =
\langle +\infty|-\infty\rangle_{\lambda}
\end{equation}
in (3.19). This relation shows that the ground state energy is
independent of the parameter $\lambda$;\ in particular one can choose
$\lambda=0$ in evaluating the ground state energy, which leads to the gauge
condition (3.4) without Gribov complications.

In the path integral (3.11), one may impose {\em periodic}
boundary conditions in time $t$ on all the integration variables
and let the time interval $\rightarrow \infty$ later so
that BRST transformation (3.15) be consistent with the boundary
conditions.

The analysis in this sub-section is general but formal. In the next
sub-section we comment on an operator Hamiltonian formalism and BRST
cohomology.\\

\noindent
{\bf 3.3, BRST COHOMOLOGY}\\
We start with the BRST invariant effective Lagrangian
\begin{eqnarray}
{\cal L}_{eff} &=& {\cal L} + {\cal L}_{g}\nonumber\\
         &=&
\frac{1}{2}\{[\dot{X}^{\omega}(t)+g\xi(t)^{\omega}Y^{\omega}(t)]^{2} +
[\dot{Y}^{\omega}(t) -
g\xi^{\omega}(t)X^{\omega}(t)]^{2} + [\dot{Z}^{\omega}(t) -
\xi^{\omega}(t)]^{2}\}\nonumber\\
&&- U[(X^{\omega}(t))^{2} +(Y^{\omega}(t))^{2}]\nonumber\\
&&-\alpha \dot{B}(t) \xi^{\omega}(t) + B(t)(Z^{\omega}(t) - \lambda
X^{\omega}(t))
+ \alpha i \dot{\bar{c}}(t)\dot{c}(t)\nonumber\\
&&-i\bar{c}(t)[ 1 + g\lambda Y^{\omega}(t)]c(t)
+\frac{\alpha}{2}B(t)^{2}
\end{eqnarray}
obtained from (3.1) and (3.18). A justification of (3.28), in particular its
treatment of Gribov-type copies, rests on the path integral representation
(3.11). In the following we suppress the suffix $\omega$, which emphasizes that
the BRST transformation is local in $\omega$-space.

One can construct a Hamiltonian from (3.28) in a standard manner as
\begin{eqnarray}
H&=& \frac{1}{2}[P_{X}^{2} +P_{Y}^{2} +P_{Z}^{2}] + U(X^{2} +Y^{2})\nonumber\\
  && + \xi G - B(Z-\lambda X) + i\frac{1}{\alpha}p_{c}p_{\bar{c}}
     + i\bar{c}( 1+\lambda gY)c - \frac{1}{2}\alpha B^{2}
\end{eqnarray}
and the Gauss operator is given by
\begin{equation}
G \equiv g(XP_{Y} - YP_{X}) + P_{Z}
\end{equation}
We note that $(p_{\bar{c}})^{\dagger} = - p_{\bar{c}}$   since  $p_{\bar{c}}=
i\alpha \dot{c}$  and  $p_{c}= -i\alpha\dot{\bar{c}}$.

The BRST charge is obtained from ${\cal L}_{eff}$ (3.28) via the
Noether current as
\begin{equation}
Q = cG - ip_{\bar{c}}B
\end{equation}
The BRST charge $Q$ is hermitian $Q^{\dagger} = Q$ and nil-potent
\begin{equation}
\{Q, Q\}_{+} = 0
\end{equation}
by noting $\{c, c\}_{+} = \{p_{\bar{c}}, p_{\bar{c}}\}_{+}= \{p_{\bar{c}},
c\}_{+} =0$. The BRST transformation (3.15) is generated  by $Q$, for example,
\begin{eqnarray}
e^{-\theta Q}X(t)e^{\theta Q} &=& X(t) - [\theta Q, X(t)]\nonumber\\
            &=& X(t) - i\theta g c(t)Y(t),\nonumber\\
e^{-\theta Q}\bar{c}(t)e^{\theta Q} &=& \bar{c}(t) - [\theta Q,
\bar{c}(t)]\nonumber\\
            &=& \bar{c}(t) + \theta B(t)
\end{eqnarray}
by noting $\theta^{2} = 0$.

The Hamiltonian in (3.29) is rewritten by using the BRST charge as
\begin{equation}
H = H_{0} + i\{ Q, \xi p_{c}\}_{+} + \{ Q, \bar{c}(Z - \lambda X\}_{+} +
\frac{\alpha}{2}\{ Q, B\bar{c}\}_{+}
\end{equation}
with
\begin{equation}
H_{0}\equiv \frac{1}{2}[  P_{X}^{2} + P_{Y}^{2} + P_{Z}^{2}] +
            U( X^{2} + Y^{2} )
\end{equation}

The  physical state $\Psi$ is defined as an element of BRST cohomology
\begin{equation}
\Psi \in \ Ker\ Q/ Im\ Q
\end{equation}
namely
\begin{equation}
Q\Psi = 0
\end{equation}
but $\Psi$ is {\em not} written in a form $\Psi = Q\Phi$ with a non-vanishing
$\Phi$.

The time development of $\Psi$ is dictated by Schroedinger equation
\begin{equation}
i\frac{\partial}{\partial t}\Psi (t) = H\Psi (t)
\end{equation}
and thus
\begin{eqnarray}
\Psi (\delta t) &=& e^{-iH\delta t}\Psi (0)\nonumber\\
&=& \Psi (0) - i\delta tH\Psi (0)\nonumber\\
&=&\Psi (0) -i\delta tH_{0}\Psi (0)\nonumber\\
&& -i\delta tQ\{ i\xi p_{c} + \bar{c}( Z- \lambda X) +
\frac{\alpha}{2}B\bar{c}\}\Psi (0)\nonumber\\
&\simeq& \Psi (0) -i\delta tH_{0}\Psi (0)
\end{eqnarray}
in the sense of BRST cohomology by noting (3.34) and $Q\Psi (0) = 0$. Note that
the Hamiltonian is BRST invariant
\begin{equation}
[ Q, H_{0}] = [ Q, H ] = 0
\end{equation}
If one solves the time independent Schroedinger equation
\begin{equation}
H_{0}\Psi (0) = E\Psi (0)
\end{equation}
with $Q\Psi (0) = 0$, one obtains
\begin{eqnarray}
\Psi (0)^{\dagger}e^{-iHt}\Psi (0) &=& e^{-iEt}\Psi (0)^{\dagger}\Psi
(0)\nonumber\\
&=& e^{-iEt}
\end{eqnarray}
by noting $\Psi (0)^{\dagger}Q = 0$. The eigen-value equation (3.41) is gauge
independent and thus $E$ is formally gauge independent.

In fact, a more detailed analysis[34] confirms that the BRST cohomology
reproduces
the result of canonical analysis in Ref.[32]. Namely, one can show that
the physical spectrum of the suluble model with the harmonic potential
\begin{equation}
U( X^{2} +Y^{2} ) = \frac{\omega^{2}}{2}( X^{2} +Y^{2} )
\end{equation}
is given by
\begin{equation}
H = \omega(n_{1} + n_{2} +1) + \frac{g^{2}}{2}(n_{1} - n_{2})^{2}
\end{equation}
where $n_{1}$ and $n_{2}$ are non-negative integers. The first term in this
formula stands for the spectrum of a two-dimensional
harmonic oscillator, and the second term stands for $\frac{1}{2} P_{Z}^{2}$
replaced by
$\frac{g^{2}}{2}L_{Z}^{2}$ by using the Gauss law operator in (3.30).
One can safely take the limit
$\alpha \rightarrow 0$  in the physical sector, though unphysical excitations
acquire
infinite excitation energy in this limit just like unphysical excitations in
gauge theory defined by $R_{\xi}$-gauge[33].

It has been shown in [32] that the correction terms arising from operator
ordering plays a crucial role in the evaluation of perturbative corrections to
ground state energy in Lagrangian path integral
formula. This problem is often treated casually in conventional perturbative
calculations;\ a general belief ( and  hope) is that Lorentz invariance and
BRST invariance somehow takes care of the operator ordering problem. It is
shown[34]
 that BRST invariance and $T^{\star}$-product prescription reproduce the
correct result of Ref.[32] provided that one uses a canonically well-defined
gauge such as $R_{\xi}$-gauge with $\alpha \neq 0$ in (3.18). This check is
important to establish the equivalence of (3.11) to the path integral formula
in Ref.[32].  If one starts with $\alpha = 0$ from the on-set, one needs
correction terms calculated in Ref.[32].

To be more explicit, what we want to evaluate is eq.(3.11), namely
\begin{equation}
\langle +\infty|-\infty\rangle = \frac{1}{\tilde{N}}{\int}d\mu
\ exp\{iS(X^{\omega},Y^{\omega},Z^{\omega},\xi^{\omega}) +i {\int}{\cal
L}_{g}dt\}
\end{equation}
We define the path integral for a sufficiently large time interval
\begin{equation}
t\in [ T/2, -T/2 ]
\end{equation}
and let $T \rightarrow \infty$ later.

The exact ground state energy  is given by eq.(3.44) as
\begin{equation}
E = \omega
\end{equation}
Namely, we have no correction depending on the gauge parameter $\lambda$ and
the coupling constant $g$. As was already shown in (3.27), the absence of
$\lambda$ dependence is a result of BRST symmetry. This property is thus more
general and, in fact, it holds for all the energy spectrum of physical states;\
this can be shown by using the Schwinger's action principle [35] and the
definition of physical states in (3.36). The perturbative check of
$\lambda$-independence or Slavnov-Taylor identities in general is carried out
in the standard manner.
On the other hand, the absence of $g$- dependence is an effect of more
dynamical origin and it can be confirmed by an explicit calculation[34]:
The ground state energy is then obtained from
\begin{eqnarray}
\langle +\infty|-\infty\rangle &=& \lim_{T\rightarrow \infty} \langle
0|e^{-iH[T/2 - (-T/2)]}|0\rangle \nonumber\\
&=& \lim_{T\rightarrow \infty} const \times e^{-i\omega T}
\end{eqnarray}
which is justified for $T = -iT_{E}$ and $T_{E} \rightarrow \infty$ in
Euclidean theory. We thus obtain the ground state energy
$E = \omega$ to be consistent with (3.47).

\section{Discussion and conclusion}

The BRST symmetry plays a central role in modern gauge theory,
and the BRST invariant path integral can be formulated by summing over all the
Gribov-type copies in a very specific manner provided that
the crucial correspondence in (2.11) or (3.17) is globally single valued[28].
This criterion is satisfied by the soluble gauge model proposed in Ref.[32],
and it is encouraging that the BRST invariant prescription is in accord with
the canonical analysis of the soluble gauge model in Ref.\cite{32}.
The detailed explicit analysis in Ref.\cite{32} and the somewhat formal BRST
analysis are complementary to each other.
In Ref.\cite{32}, the problem related to the so-called Gribov horizon , in
particular the possible singularity associated with it, has been analyzed in
greater detail;\ this is crucial for the analysis of more general situation. On
the other hand, an advantage of the BRST analysis is that one can clearly see
the gauge independence of physical quantities such as the energy spectrum as a
result of BRST identity.

The BRST approach allows a transparent treatment of general class of gauge
conditions implemented by (3.18). This gauge condition with $\alpha \neq 0$
renders the canonical structure better-defined, and it allows simpler
perturbative treatments of the problems such as the corrections to the ground
state energy. The BRST  analysis in Ref.[34] vis-a-vis the explicit canonical
analysis in Ref.\cite{32} may provide a justification of conventional covariant
perturbation theory in gauge theory, which is based on Lorentz invariance( or
$T^{\star}$-product ) and BRST invariance without the operator ordering terms.

Motivated by the observation in Ref.[32] to the effect  that the Gribov
horizons are not really singular in quantum mechanical sense, which is in
accord
with our path integral in (3.21), we would like to make a
speculative comment on the role of Gribov copies in QCD.
Some of the non-perturbative effects such as quark confinement and hadron
spectrum may be analyzed at least qualitatively in the $1/N$ expansion sheme,
for
example, without referring to non-trivial topological structures of gauge
fields[36]. This scheme is based on a sum of an infinite number of Feynman
diagrams.
 This diagramatic approach
or  an analytical treatment equivalent to it in the Feynman-type gauge deals
with topologically trivial gauge fields but it may still suffer from the Gribov
copies, as is suggested by the analysis in Ref.[27]:
If one assumes that the vacuum is unique in this case as is the case in
$L^{2}$-space, the global single-valuedness in (2.11) in the context of path
integral (2.10) will be preserved for infinitesimally small fields $A_{\mu}$.
By a continuity argument, a smooth deformation of $A_{\mu}$ in $L^{2}$-space (
or its extension as explained in Section 2) will presumably keep the integral
(2.10) unchanged. ( A kind of topological invariant).  If this argument is
valid, the formal path integral formula (2.5) will
provide a basis for the analysis of some non-perturbative aspects of QCD.

On the other hand, the Gribov problem may also suggest the presence of some
field configurations which do not satisfy any given gauge condition in four
dimensional non-Abelian gauge theory [25]. For example, one may not be able to
find any gauge parameter $\omega (x)$ which satisfies
\begin{equation}
\partial^{\mu}A_{\mu}^{\omega}(x) = 0
\end{equation}
for some fields $A_{\mu}$. Although the measure of such field configurations in
path integral is not known, the presence of such filed configurations would
certainly modify the asymptotic correspondence in
(2.11).  In the context of BRST symmetry, the Gribov problem may then induce
complicated phenomena such as the dynamical instability
of BRST symmetry[37]. If the dynamical instability of BRST symmetry should take
place, the relation corresponding to (3.27), which is a
result of the BRST invariance of the vacuum, would no longer be derived.  In
the framework of path integral, this failure of (3.27) would be recognized as
the failure of the expansion (3.19) since the normalization factor $N$ in
(3.21) would generally depend on not
only field variables but also $\lambda$ if the global single-valuedness in
(3.17) should be violated.

In Ref.(37), a {\em  necessary} condition for the spontaneous breakdown of
BRST symmetry was formulated as
\begin{equation}
Tr( e^{i\pi Q_{c}}) = 0
\end{equation}
with $Q_{c}$ the ``ghost number `` operator. This relation is analogous to the
Witten index in conventional supersymmetric theories[38]. It would be
interesting to analyze the $1/N$ expansion of QCD from a view point of the
above index condition (4.2); in other words, if one can prove that the index is
non-zero $Tr( e^{i\pi Q_{c}})\neq  0 $ for the field configurations relevant
for the $1/N$ expansion, our speculation on the possible BRST symmetric
treatment of some non-perturbative aspects of QCD will be justified.

Finally, we note that the lattice gauge theory [39], which is based on
compactified field variables, is expected to change the scope and character of
the Gribov problem completely.

\begin{figure}
\epsfbox{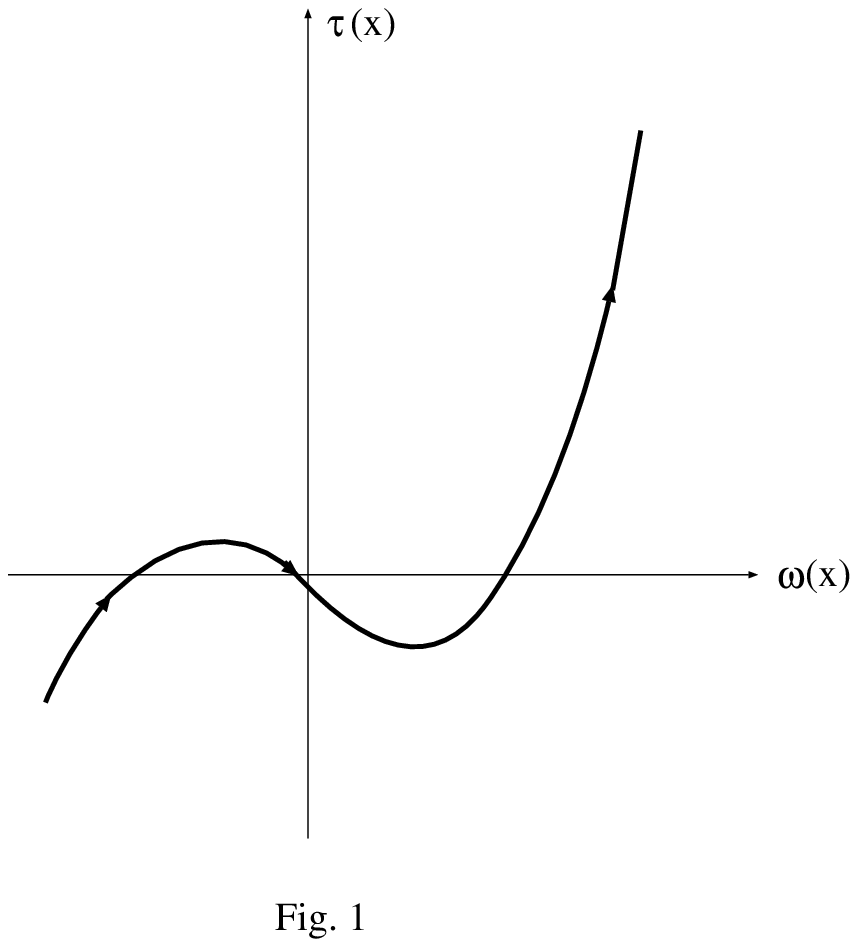}
\caption{ A schematic representation of eq. (2.11) for fixed
$A_{\mu}$.}
\end{figure}

\end{document}